\documentclass[12pt,preprint]{aastex}

\slugcomment{Accepted by \it{The Astrophysical Journal}}
\shorttitle{X-ray Emission from PN Hb 5}
\shortauthors{Montez et al.}
\begin{document}
\title{Serendipitous XMM-Newton Detection of X-ray Emission from the Bipolar Planetary Nebula Hb 5}
\author{Rodolfo Montez Jr. and Joel H. Kastner} 
\affil{Chester F. Carlson Center for Imaging Science, Rochester Institute of Technology,
    Rochester, NY 14623}
\email{rodolfo.montez.jr@gmail.com,jhk@cis.rit.edu}

\author{Bruce Balick}
\affil{Department of Astronomy, University of Washington, Seattle, WA 98195-1580}
\email{balick@astro.washington.edu}

\and 

\author{Adam Frank}
\affil{Department of Physics and Astronomy and C. E. K. Mees Observatory, University of Rochester, Rochester, NY 14627-0171}
\email{afrank@pas.rochester.edu}

\begin{abstract}
We report the serendipitous detection by the XMM-Newton X-ray Observatory 
of an X-ray source at the position of the Type I (He- and N-rich) 
bipolar planetary nebula Hb 5.  The Hb 5 X-ray source appears marginally resolved.  
While the small number of total counts ($\sim 170$) and significant off-axis angle of the 
X-ray source ($\sim 7.8^{\prime}$) precludes a definitive spatial analysis, 
the morphology of the X-ray emission appears to trace 
the brightest features seen in optical images of Hb 5.  
The X-ray spectrum is indicative of a thermal plasma at a temperature between 
2.4 and 3.7 MK and appears to display strong Neon emission.  The inferred X-ray luminosity is $L_X = 1.5 \times 10^{32}\textrm{ ergs s}^{-1}$.  
These results suggest that the detected X-ray emission 
is dominated by shock-heated gas in the bipolar nebula, although we cannot 
rule out the presence of a point-like component at the position of the 
central star.  The implications for and correspondence with current models 
of shock-heated gas in planetary nebulae is discussed.
\end{abstract}
\section{Introduction}

The shaping and evolution of planetary nebulae (PNe) has been an active 
field of study for a few decades and is continually spurred on by new 
observations that do not fit the existing theoretical models \citep[and references therein]{bal02}.  
Of particular note are the bipolar PNe, with 
their large lobes and narrow waists, many showing residual evidence of 
a spherical AGB wind in the form of nested rings centered on the PN central star \citep{cor04}.  
The implied transition from seemingly spherical mass loss to profoundly
bipolar outflow remains an unsolved mystery.    
Possible shaping and evolutionary mechanisms have been developed ranging from the interacting 
stellar winds model \citep{kwo78}, wherein a hot fast stellar wind shocks a slower 
moving AGB wind, to collimated flows \citep{sok04,aka08} possibly launched and shaped by 
a magnetocentrifugal force fueled by an AGB dynamo \citep{bla01}.
 
In all of these PN shaping models the presence of gas heated to X-ray emitting temperatures is 
predicted.  The advent of high-spatial-resolution, X-ray observatories 
(Chandra and XMM-Newton) facilitated the discovery and scrutiny of this 
shock-heated gas \citep[and references therein]{gue05,kas08} and the results were surprising.  The observed 
X-ray temperatures of the shocked plasma have been much lower, $1 - 3 \times 10^6$ K, 
than predicted, $> 10^{7}$ K \citep{stu06,kas08}, spawning a variety of potential 
solutions \citep[see][for a review]{sok03}. 
\citet{aka06,aka07,aka08} 
and \citet{ste08} have since produced more detailed models in an attempt 
to resolve the temperature discrepancy while adequately reproducing other
observed quantities.  
\citet{ste08} investigate the role played by thermal conduction in establishing the 
physical properties of the X-ray emitting gas, while \citet{aka06,aka07} consider 
the possibility that the X-ray emitting gas arose from a slower, post-AGB wind 
($\sim 500$ km/s), and \citet{aka08} present numerical simulations that suggest 
the observed X-ray properties could be explained by jet-wind interaction.
These newer models describing PN X-ray emission should, in turn, provide feedback to generalized
models describing the origin of the flows shaping PNe, which, taken together, will 
provide a complete theoretical description of the shaping and evolution of PNe. 

The identification of X-ray sources within bipolar PNe should provide unique constraints 
on this new generation of PN shaping and evolution models. However, thus far 
such detections have been few and far between \citep{gru06,kas01,kas03,sah03}. 
During an archival program to identify serendipitous X-ray sources associated with PNe,  
we established that the XMM-Newton Serendipitious Survey\footnote{See \citet{wat07} 
for information on the XMM-Newton Serendipitous Survey catalog.} includes an X-ray source, 
2XMM J174756.2-295937, that is spatially coincident, within errors ($\sim\pm 2^{\prime\prime}$), 
with \object[PN HB 5]{Hubble 5} (Hb 5), a large, narrow-waist bipolar planetary nebula with 
a bright, compact, core \citep{pot07}.  
\citet{qui07} classify Hb 5 as Peimbert Type I PN, i.e., a He- and N-rich PN 
descended from a relatively massive progenitor star.  Hb 5 is a high-excitation PN and, 
in particular, is one of the few PNe showing the 7.652 $\mu$m Ne VI line \citep{pot07}.  
A photodissociation region immediately surrounding the nebula may account for 
the presence of molecular hydrogen emission \citep{ber05,pot07}.
The distance to Hb 5 is highly uncertain. \citet{ric04} derive a distance 
of 1.4 kpc $\pm$ 0.3 kpc from 3D photoionization modeling, while \citet{pot07} 
argue for a much larger distance, ranging from 3 to 7 kpc.  \citet{cor93} find 
a polar expansion velocity of $\sim250$ km s$^{-1}$, while \citet{pis00} reported 
a faster, possibly collimated, wind of 400 km s$^{-1}$ near the center of Hb 5.

\section{Analysis}
\subsection{Data}

Hb 5 lies within the field of an XMM-Newton observation of the pulsar PSR J1747-2958 
and pulsar wind nebula (MOUSE NEBULA).  This observation (ObsID 0152920101) was 
performed during XMM Revolution 607 on 2003 April 02.  
The European Photon Imaging Camera (EPIC) detector arrays 
pn, MOS1, and MOS2 were operated in Full-Frame Mode, with 
the thick filter, for total exposures of 45.1, 51.3, and 
51.3 ks, respectively.  
We reprocessed the XMM-Newton Observation Data Files 
using the XMM-Newton Science Analysis Sofware (SAS) package version 7.1.0 
with the calibration files available in Current Calibration File 
Release 241 (XMM-CCF-REL-241).  We filtered out high background 
periods and bad events from all observations using standard filters 
for imaging mode observations.  The resulting net exposure times are 44.1, 
50.6, and 50.8 ks, in the pn, MOS1, and MOS2 arrays, respectively.  

From the source and background regions used in the spectral extraction 
of the X-ray source at the position of Hb 5 (described in Section 2.3), 
we determined that the net source and background count rates in the 
energy range 0.2-2.0 keV are
$1.9 \pm 0.4 \times 10^{-3} \textrm{ cnts s}^{-1}$ in pn, 
$8.0 \pm 1.8 \times 10^{-4} \textrm{ cnts s}^{-1}$ in MOS1,
and $8.7 \pm 1.9 \times 10^{-4} \textrm{ cnts s}^{-1}$ in MOS2.  
We merged the three observations using the SAS task \textit{merge}.  
Using this merged event list, we generated a soft band image for the energy 
range 0.2-2.0 keV, binned to 5$^{\prime\prime}$ pixels (Figure \ref{fig1}a).  

\subsection{Spatial Analysis}

\begin{figure}
\epsscale{.75}
\plotone{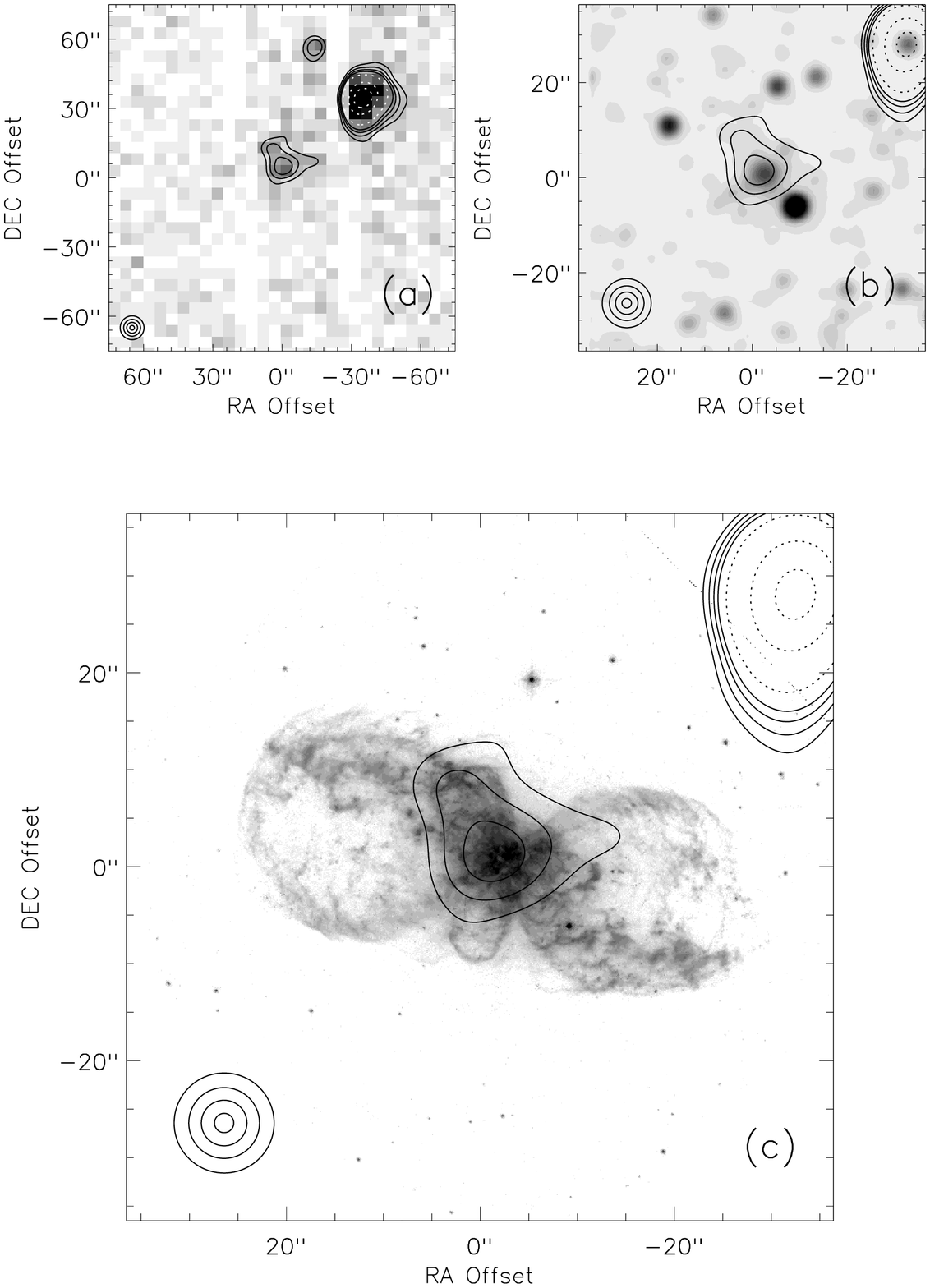}
\caption{XMM imaging of the Hb 5 field.  Smoothed X-ray contours at 9, 11, 
13, and 15 counts (solid contours) and at 20, 40, and 60 counts (dotted contours) 
are overlaid upon the (a) unsmoothed XMM X-ray image, (b) 2MASS J-band image, and (c) HST F658N narrowband image ($H\alpha + [NII]$).  In the lower left corner of each panel, we display 
the 25\%, 50\%, 75\% and 90\% levels of the normalized Gaussian filter used to smooth the X-ray image.\label{fig1}}
\end{figure}

Figure \ref{fig1}a indicates that the X-ray emission at the position of Hb 5
is marginally resolved.  Hb 5 is located $\sim7.8^{\prime}$ off-axis in the 
XMM observation, and the coincident X-ray source displays peak emission around 1 keV.  
This off-axis angle and peak emission are within the range of application of the 
XMM/EPIC PSF model described 
in \citet{ghi02}, i.e., a King profile whose core radius and slope depend on 
energy and off-axis angle.  Specifically, we expect a PSF core radius of 
4.71$^{\prime\prime}$ and slope of -1.44. This suggests that using a normalized 
Gaussian smoothing filter with a FWHM of 7.5$^{\prime\prime}$ will reduce the 
loss of information, while aiding our interpretation of the detected X-ray emission. 
In Figure \ref{fig1}, the smoothed X-ray image contours are overlaid onto the 
unsmoothed X-ray image (Figure \ref{fig1}a), a 2MASS J-band image (Figure \ref{fig1}b), 
and a HST WFPC2 F658N ($H\alpha + [NII]$) narrowband image (Figure \ref{fig1}c).  The 
image registration was performed by the IDL Astro 
Library\footnote{http://idlastro.gsfc.nasa.gov} task \textit{hastrom}.  
A slight eastward shift ($\sim 0.45^{\prime\prime}$) of the HST image was required 
to align the stars appearing in both the 2MASS J-band and chip 3 of the HST WFPC2 F658N fields. 
Figures \ref{fig1}b and \ref{fig1}c demonstrate that the peak of the X-ray emission coincides 
with the bright core regions of Hb 5 seen in the 2MASS and HST images, respectively.  
The apparent extension of the X-ray emission lies along the brightest features observed 
in the $H\alpha + [NII]$ image. 

\subsection{Spectral Analysis}

The spatial region available for spectral extraction within the XMM image 
of the Hb 5 field is severely constrained by several unavoidable artifacts.  
Immediately surrounding the region of 
interest are chip gaps (in all three EPIC arrays), a bright nearby soft source, 
and a readout streak from the nearby soft source.  
Guided by the optical HST position of Hb 5 along with 
the evident soft X-ray emission, we 
selected an extraction region that optimized the signal in the spectrum 
while avoiding these artifacts.  An annulus surrounding the 
source region and omitting nearby sources was used to estimate the background.
We used the SAS tasks \textit{rmfgen} and \textit{arfgen} to generate source-specific 
response matrices and effective area curves; the latter accounts for off-axis 
vinetting ($\sim30\%$ loss at $7.8^{\prime}$, \citet{ehl08}).

\begin{figure}
\epsscale{.80}
\plotone{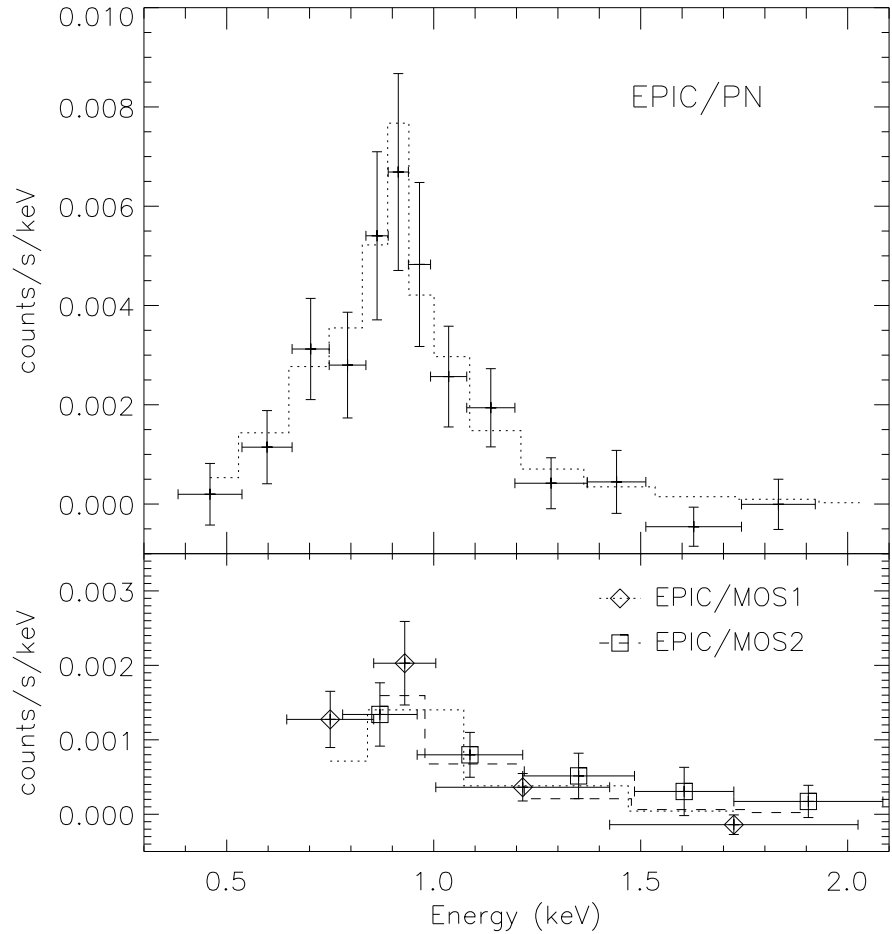}
\caption{X-ray spectra of the Hb 5 source as obtained from the three 
EPIC CCD detector arrays (crosses, diamonds, and squares; pn: top panel; 
MOS: bottom panel) overlaid with the result of the best 
simultaneous fit to a model consisting of a thermal plasma (\textit{vmekal}) 
and intervening absorption (\textit{wabs}).\label{fig2}}
\end{figure}

The resulting spectra from the three EPIC CCD detector arrays were simultaneously fit 
in XSPEC (version 12.3.1; \citet{arn96}) with a variable abundance thermal plasma 
model, \textit{vmekal} \citep{mew85,mew86,kaa92,lie95}, and 
intervening absorption, \textit{wabs} \citep{mor83}.  We used the extinction value 
$C = 1.60$ at H$\beta$ listed in \citet{pot07} to fix the column density at
$N_H = 6.0 \times 10^{21} cm^{-2}$ and -- noting the apparent presence of a strong 
spectral feature at $\sim 1.0$ keV that is likely due to Ne IX and Ne X lines -- left 
as free parameters the gas temperature and the abundance of Ne.  We fixed all 
other abundances at their solar values\citep{and89}.  Although fixing the abundance of Ne to its 
solar value results in a global fit that is acceptable, this fit is poor at the spectral 
feature near 1.0 keV.  There is a modest improvement in the fit when we allow
the abundance of Ne to be a free parameter.  This latter fit is presented in Figure \ref{fig2}.  
The temperature is found to lie in the range 2.4-3.7 MK ($90\%$ confidence range) 
and the Ne abundance factor lies in the range 1.8-6.9 ($90\%$ confidence range) 
relative to the solar abundance given by \citet{and89}.  
The best model fit indicates an observed 0.2-2.0 keV flux of
$7.9 \times 10^{-15} \textrm{ ergs cm}^{-2}\textrm{ s}^{-1}$, and an 
unabsorbed flux of $1.2 \times 10^{-13}  \textrm{ ergs cm}^{-2}\textrm{ s}^{-1}$. 
We assume a distance of 3.2 kpc after \citet{pot07} and find $L_X = 1.5 \times 
10^{32} (D/(3.2\textrm{ kpc}))^{2}\textrm{ ergs s}^{-1}$.  
  
\section{Discussion}

Hb 5 is the fourth bipolar PNe -- after NGC 7026 \citep{gru06}, 
the transitional object NGC 7027 \citep{kas01}, and the possibly symbiotic 
system Mz 3 \citep{kas03} -- with suspected diffuse X-ray emission and only 
the second Type I PN known to display X-ray emission.
The other Type I PN with evidence for shocked X-ray emission is the $[$WC$]$ 
PN NGC 5315, which was also discovered serendipitously, far off-axis, by the 
Chandra X-ray Observatory \citep{kas08}.  Indeed, the X-ray emission regions  
detected in Hb 5 and NGC 5315 display similar characteristics,
despite the fact that one is associated with a high excitation, bipolar PN and the 
other with a compact $[$WC$]$ PN.  Specifically, these two PNe have 
similar plasma temperatures (within errors) and similar X-ray luminosities (though 
the $L_X$ values are very uncertain, due to distance uncertainties).
In addition, both appear to show strong X-ray Ne lines.
\citet{pot07} find an Ne abundance factor of 2.2 relative to solar in the infrared 
spectrum of Hb 5, similar to the (uncertain) overabundance tentatively
determined from the X-ray spectrum.

The bipolar PNe detected thus 
far have closed bipolar lobes, while many of the bipolar PNe observed 
but not detected have open bipolar lobes \citep{gru06}.
The role of open and closed lobes is an important factor that may determine
key characteristics regarding the presence or absence of hot gas due to
wind shocks; open lobes allow the gas to expand and, thus,
quickly cool, whereas closed lobes evidently contain the
gas, and cooling through heat conduction at the nebular-hot
gas boundary may dominate \citep{ste08}.  In this regard, it appears significant 
that the extended X-ray emission in Hb 5 corresponds closely to the brighter
regions seen in the HST image, suggesting the X-rays are coming from
shocked material within the most tightly confined regions of the lobes.

Recent simulations of axisymmetric jets expanding into a 
spherical wind by \citet{aka08} reproduced optical features seen in 
some bipolar PNe. This study focused on Mz 3 and M 1-92, but 
included simulations with generic PNe parameters. Some of these key optical 
features are also seen in Hb 5, i.e., an equatorial dense region and 
bright polar rim.  The evidence of a fast, possibly collimated, wind 
reported by \citet{pis00} suggests that the extended X-ray emission observed 
in Hb 5 can likely be traced to a collimated fast wind \citep{aka06,aka07}.  
The point-symmetric features of the nebula and the extension in the 
X-ray emission support the possibility of an X-ray jet or shocks from such a 
collimated flow.  The spatial coincidence with the central region of Hb 5 
suggests that some X-rays may originate from the core and/or the central star 
itself.  However, caution is warranted, given the small number of counts 
detected due to vingetting at the large off-axis angle of Hb 5 and the large background 
due to nearby bright sources convolved with the XMM PSF.

A deep, targeted X-ray observation of Hb 5 might bring our 
Ne abundance factor into agreement with the infrared-determined value and 
would decrease the uncertainties in Ne abundance by allowing us to fit the 
abundances of additional elements.  Also, in the archival XMM-Newton observation 
of Hb 5, small but significant (termination) portions 
of the bipolar lobes lie in the detector chip gap and the readout streak 
of the nearby source.
Hence, a targeted Chandra X-ray Observatory X-ray observation of Hb 5 would 
provide the higher spatial resolution imagery that might enhance the 
role this PN plays in our 
understanding of the shaping and evolution of planetary nebulae.  

\section{Conclusions}

Our spatial and spectral analysis of a moderately resolved, off-axis, low count 
XMM-Newton detection of X-ray emission at the location of the bipolar PN Hb 5 
leads us to conclude that the emission originates with Hb 5.  
The X-ray emission is indicative of a thermal 
plasma at 2.4-3.7 MK with X-ray luminosity of $1.5 \times 10^{32} \textrm{ erg s}^{-1}$.  
Comparison with images of infrared and 
optical emission shows that the peak of the X-ray emission corresponds closely 
with the bright core of Hb 5 and the extension of the X-ray emission lies along bright 
extensions in the optical image.   
The X-ray temperature and luminosity, as compared to PNe known to exhibit
shock-induced X-ray emission \citep{kas08}, further suggest that 
the emission arises from shocks in the nebula, as opposed to the harder and less 
luminous X-ray emission associated with central stars (e.g. Mz 3; \citet{kas03}).
As only the fourth bipolar PNe with detected X-ray emission, 
Hb 5 may play an important role in the evolving study of the shaping and evolution 
of PNe. This PN therefore merits future targeted, deep, high spatial resolution 
X-ray observation. 

\acknowledgments
This research was supported by NASA Astrophysics Data Analysis Program award NNX08AJ65G to RIT.

{\it Facilities:} \facility{XMM (EPIC)}, \facility{HST (WFPC2), \facility{CTIO:2MASS}}




\clearpage



\clearpage

\end{document}